\documentclass[prb,preprint]{revtex4-1} 

\usepackage{amsmath}  
\usepackage{amsfonts} 
\usepackage{graphicx} 

\begin{document}


\title{Spin contribution to the perihelion precession in binary systems like OJ287: higher order corrections.}


\author{Carlos Mar\'in}
\email{cmarin@usfq.edu.ec}
\affiliation{Department of Physics, Universidad San Francisco de Quito, Diego de Robles y V\'ia Interoce\'anica }

\author{Jorge Poveda }
\email{jorge.okuden@gmail.com} 
\affiliation{Department of Physics, Universidad San Francisco de Quito, Diego de Robles y V\'ia Interoce\'anica} 



\date{\today}
\begin{abstract}{Higher order corrections are obtained for the perihelion precession in binary systems like OJ287, Sagittarius A*-S2 and H1821+643 using both the Schwarzschild metric and the Kerr metric to take into account the spin effect. The corrections are performed considering the third root of the motion equation and developing the expansion in  terms of $\epsilon \equiv r_s/\left(a(1-e^2)\right)$ and $\epsilon^{*} \equiv \left(1- \frac{2 \alpha E'}{cJ}\right) \epsilon $.The results are compared  with those obtained in a previous paper.} 

\end{abstract}


\maketitle 
\section{Introduction}
\label{intro}

\begin{figure}
\begin{centering}
\includegraphics{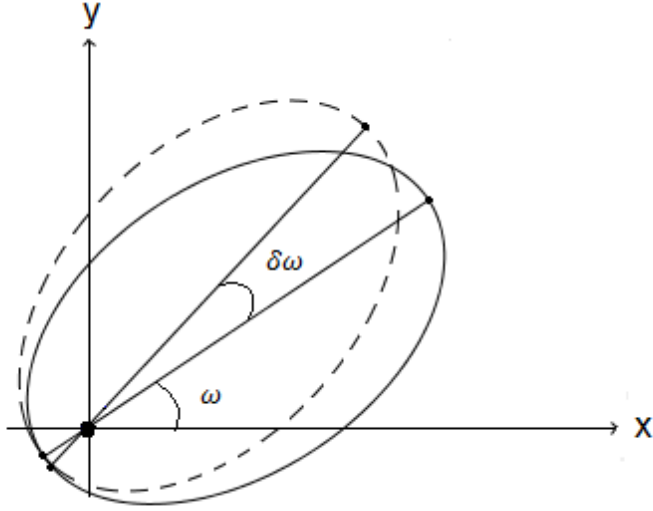}
\par\end{centering}
\caption{Perihelion precession. $\omega$ is the initial inclination of the orbit and $\delta\omega$ is the angle of precession.\label{fig:avance}}
\end{figure}

In a previous paper \cite{MarinPoveda},  higher order corrections (up to n-th order) were obtained for the perihelion precession (see figure  (\ref{fig:avance})) in binary systems like OJ287, Sagittarius A*-S2 and H1821+643  using the Schwarzschild metric and complex integration. The corrections were performed considering the third root of the motion equation and developing the expansion in  terms of $\epsilon \equiv r_s/\left(a(1-e^2)\right)$, where $r_{s}$  is the Schwarzschild radius.The results were compared  with other expansions that appear in the literature giving corrections to second and third order \cite{Fokas,Tyler,Rosales,Biesel,DEliseo,Scharf,Do_Nhat}. In this paper we will consider both Schwarzschild and Kerr metrics to take into account the spin effect and developing the expansion in terms of both $\epsilon \equiv r_s/\left(a(1-e^2)\right)$ and  $\epsilon^{*} \equiv \left(1- \frac{2 \alpha E'}{cJ}\right) \epsilon $, where $E'$ and $J$ are the energy and angular momentum per unit mass and $\alpha$ is a factor proportional to the black hole spin.

Kerr's black holes are very interesting because most of the black holes in the universe probably have a rotational movement (spin) and one of the most important consequences of that spin is that spacetime is dragged around the rotating black hole leading to an effect that is known as "frame dragging". In the case of a Schwarzschild black hole, a precession occurs which is the rotation of the elliptical orbit in the fixed plane of that orbit. For Kerr's solution, the drag of the reference frame introduces an additional precession of the plane of the orbit around the axis of rotation of the black hole in the same direction of rotation of said black hole. To facilitate the calculations we will assume that the axis of rotation of the Kerr black hole is perpendicular to the plane of the orbit.

\section{Spin contribution to the perihelion precession}

\medskip{}

The Schwarzschild metric describes a body with spherical symmetry, but without electric charge and rotational movement.  If we take into account that many black holes that have been found have rotation on their own axis, as for example the binary  system OJ287, one might think that this rotation should influence the advance of the perihelion of the elliptical orbits. The rotation around its own axis is given by the angular momentum of spin $ S_ {z} $ of the massive body M.

To include the spin in the calculation, you can use the Kerr metric. The Kerr metric is a solution to the field equations in vacuum for a body of mass M that rotates on its own axis with an angular momentum $ S_ {z} $. The Kerr metric is \cite{Misner,Ryder,Hobson,Chandrasekhar,tHooft,Ludvigsen,Marin}:
\\
\begin{eqnarray}
\left(ds\right)^{2}=c^{2}(d\tau)^{2}=\gamma c^{2}(dt)^{2}-\frac{r^{2}}{\Delta}(dr)^{2}-r^{2}\left(1+\frac{\alpha^{2}}{r^{2}}+\frac{r_{s}\alpha^{2}}{r^{3}}\right)(d\phi)^{2}+\frac{2r_{s}\alpha}{r}cdtd\phi,
\end{eqnarray}
\\
where $\alpha=\frac{S_{z}}{Mc}$, $\Delta=r^{2}-r_{s}r+\alpha^{2}$, with coordinates  $x^{0}=ct$, $x^{1}=r$, $ x^{2}=\theta$ and  $x^{3}=\phi$. $r_{s}=\frac{2GM}{c^{2}}$ is the Schwarzschild radius. We have taken  $\theta=\pi/2$ (equatorial plane). In this paper we only intend to observe how the spin could contribute to the perihelion precession  and if it is relevant to the calculation, for which only first order terms in $ \frac{\alpha}{r} $ will be taken into account. By doing this, the metric is reduced to:
\\
\begin{eqnarray}
\left(ds\right)^{2}=c^{2}(d\tau)^{2}=\gamma c^{2}(dt)^{2}-\frac{1}{\gamma}(dr)^{2}-r^{2}(d\phi)^{2}+\frac{2r_{s}\alpha}{r}cdtd\phi.   \label{eq:mov}
\end{eqnarray}

The arc length  $ds$ satisfies the relation $ds^{2} = g_{\mu \nu}dx^{\mu}dx^{\nu}$, then, the covariant metric tensor is:

\begin{eqnarray}
g_{\mu\nu}=\left(\begin{array}{cccc}
\gamma & 0 & 0 & \frac{r_{s}\alpha}{r}\\
0 & -\gamma^{-1} & 0 & 0\\
0 & 0 & 0 & 0\\
\frac{r_{s}\alpha}{r} & 0 & 0 & -r^{2}.
\end{array}\right)
\end{eqnarray}

The geodesic equation can be written in an alternative form using the Lagrangian 

\begin{eqnarray}
L\left(x^{\mu}, \frac{dx^{\mu}}{d\sigma}\right) = - g_{\alpha \beta}\left(x^{\mu}\right)
\frac{dx^{\alpha}}{d\sigma}\frac{dx^{\beta}}{d\sigma} 
\label{eq:Lagrangian RN}
\end{eqnarray}
\\
where $\sigma$ is a parameter of the trajectory of the particle, which is usually taken to be the proper time, $\tau$ for a massive particle. Using the Euler-Lagrange equations:

\begin{eqnarray}
\frac{\partial L}{\partial x^{\mu}}-\frac{d}{d \sigma}\left(\frac{\partial L}{\partial \left(\frac{dx^{\mu}}{d \sigma}\right)}\right)=0
\end{eqnarray}
\\
we get the geodesic equation for the particle:

\begin{eqnarray}
\frac{du_{\mu}}{d\sigma} = \frac{1}{2} \left(\partial_{\mu} g_{\alpha \beta}\right) u^{\alpha} u^{\beta}  \label{eq:geodesic}
\end{eqnarray}
where $u_{\mu} = \frac{dx_{\mu}}{d \sigma}$.

With $\sigma= \tau$, for the coordinates $ct$ ($\mu=0$) and $\phi$ ($\mu = 3$), the geodesic equation (\ref{eq:geodesic}) give us, respectively:

\begin{eqnarray}
\frac{d}{d\tau}\left(\gamma c^{2}\left(\frac{dt}{d\tau}\right)+\frac{r_{s}\alpha c}{r}\left(\frac{d\phi}{d\tau}\right)\right)=0,
\label{eq:mu0}
\end{eqnarray} 

\begin{eqnarray}
\frac{d}{d\tau}\left(r^{2}\left(\frac{d\phi}{d\tau}\right)-\frac{r_{s}\alpha c}{r}\left(\frac{dt}{d\tau}\right)\right)=0
\label{eqmu3}
\end{eqnarray}

Both of these equations define the following constants along the trajectory of the particle around the massive object:

\begin{eqnarray}
\gamma c^{2}\left(\frac{dt}{d\tau}\right)+\frac{r_{s}\alpha c}{r}\left(\frac{d\phi}{d\tau}\right)=E',
\label{eq:mu00}
\end{eqnarray} 

\begin{eqnarray}
r^{2}\left(\frac{d\phi}{d\tau}\right)-\frac{r_{s}\alpha c}{r}\left(\frac{dt}{d\tau}\right)=J
\label{eqmu33}
\end{eqnarray}
where $E'$ has units of energy per unit mass and $J$ of angular momentum per unit mass.

From these equations we get:

\begin{equation}
\label{eq:dphitau}
\frac{d\phi}{d\tau}=\frac{1}{\gamma cr^{2}}\left(\frac{r_{s}\alpha}{r}E'+\gamma cJ\right)
\end{equation}

\begin{equation}
\frac{dt}{d\tau}=\frac{1}{\gamma cr^{2}}\left(\frac{r^{2}}{c}E'-\frac{r_{s}\alpha}{r}J\right)
\end{equation}

Equation (\ref{eq:mov}) can be written as:

\begin{equation}
c^{2}=\gamma c^{2}\left(\frac{dt}{d\tau}\right)^{2}-\frac{1}{\gamma}\left(\frac{dr}{d\tau}\right)^{2}-r^{2}\left(\frac{d\phi}{d\tau}\right)^{2}+\frac{2r_{s}\alpha}{r}c\frac{dt}{d\tau}\frac{d\phi}{d\tau},
\end{equation}
\\
and replacing the values of  $\frac{dt}{d\tau}$ and  $\frac{d\phi}{d\tau}$:

\[
\gamma^{2}r^{4}c^{4}=\gamma c^{2}\left(\frac{r^{2}}{c}E'-\frac{r_{s}\alpha}{r}J\right)^{2}-\gamma c^{2}r^{4}\left(\frac{dr}{d\tau}\right)^{2}-r^{2}\left(\frac{r_{s}\alpha}{r}E'+\gamma cJ\right)^{2}\qquad\qquad
\]

\[
\qquad\qquad\qquad\qquad\qquad\qquad+\frac{2r_{s}\alpha}{r}\left(r^{2}E'-\frac{r_{s}\alpha}{r}cJ\right)\left(\frac{r_{s}\alpha}{r}E'+\gamma cJ\right).
\]

Simplifying and taking only first order terms in  $\frac{\alpha}{r}$:
\\
\[
\gamma r^{4}c^{4}=c^{2}\left(\frac{r^{4}}{c^{2}}E'^{2}-\frac{2rr_{s}\alpha}{c}JE'\right)-c^{2}r^{4}\left(\frac{dr}{d\tau}\right)^{2}-r^{2}\left(2c\frac{r_{s}\alpha}{r}JE'+\gamma c^{2}J^{2}\right)+\frac{2r_{s}\alpha}{r}\left(r^{2}cJE'\right).
\]

Introducing the value of  $\gamma$ in the left side of the last equation we have:
\\
\[
c^{2}-\frac{r_{s}c^{2}}{r}=\frac{1}{c^{2}}E'^{2}-\frac{2r_{s}\alpha}{cr^{3}}JE'-\left(\frac{dr}{d\tau}\right)^{2}-\gamma\frac{J^{2}}{r^{2}}
\]

Finally we obtain an energy conservation equation, similar to the one obtained in the case of the Schwarzschild metric, but with an additional crossed term proportional to $J E'$ :

\begin{eqnarray}
\frac{E'^{2}}{c^{2}}-c^{2}=\left(\frac{dr}{d\tau}\right)^{2}+\gamma\frac{J^{2}}{r^{2}}-\frac{r_{s}c^{2}}{r}+\frac{2r_{s}\alpha}{cr^{3}}JE'
\label{eq:conservedenergy}
\end{eqnarray}

This alllows us to define an effective potential:
\\
\begin{equation}
\widetilde{V}=\gamma\frac{J^{2}}{r^{2}}-\frac{r_{s}c^{2}}{r}+\frac{2r_{s}\alpha}{cr^{3}}JE'
\end{equation}

From equation (\ref{eq:conservedenergy}), we can get the expression for the radial kinetic energy per unit mass:
\\
\begin{eqnarray}
\left(\frac{dr}{d\tau}\right)^{2}=A+\frac{r_{s}c^{2}}{r}-\frac{J^{2}}{r^{2}}+\frac{J^{2}r_{s}}{r^{3}}-\frac{2r_{s}\alpha}{cr^{3}}JE'
\label{eq:kerr},
\end{eqnarray}
\\
where $A= \frac{E'^{2}}{c^{2}}-c^{2}$.

Recall now that since the orbit is an ellipse, there are two points in which the temporal derivative becomes zero, and they are aphelion and perihelion. For these two points we can write:

\begin{eqnarray}
A+\frac{r_{s}c^{2}}{R_{a}}-\frac{J^{2}}{R_{a}^{2}}+\frac{J^{2}r_{s}}{R_{a}^{3}}-\frac{2r_{s}\alpha}{cR_{a}^{3}}JE'=0
\label{eq:3.4-1}
\end{eqnarray}

\begin{eqnarray}
A+\frac{r_{s}c^{2}}{R_{p}}-\frac{J^{2}}{R_{p}^{2}}+\frac{J^{2}r_{s}}{R_{p}^{3}}-\frac{2r_{s}\alpha}{cR_{p}^{3}}JE'=0,
\label{eq:3.5-1}
\end{eqnarray}

where $R_{a}=a\left(1+e\right)$ and $R_{p}=a\left(1-e\right)$.

For an ellipse, the equation of motion have three real and positive roots. Two of the roots are  
 $R_{a}$ and  $R_{p}$  and the other we will call  $R'_{o}$. To obtain this root ($R'_{o}$) we use equation (\ref{eq:kerr}) :
 \\

\[
\left(\frac{dr}{d\tau}\right)^{2}\frac{r^3}{J^2}=\frac{A}{J^2}r^3+\frac{r_{s}c^{2}}{J^2}r^2-r+r_s-\frac{2r_{s}\alpha E'}{cJ}=0,
\]
and write:
\[
\frac{A}{J^2}r^3+\frac{r_{s}c^{2}}{J^2}r^2-r+r_s-\frac{2r_{s}\alpha E'}{cJ}=\frac{A}{J^2}(r-R_a)(r-R_p)(r-R'_o).
\]

The last equation can be written as:

\[
\frac{r_{s}c^{2}}{J^{2}}r^{2}-r+\left(r_{s}-\frac{2r_{s}\alpha}{cJ}E'\right)=\frac{\left|A\right|}{J^{2}}\left(R'_{o}+R_{a}+R_{p}\right)r^{2}\qquad\qquad\qquad\qquad\]

\[\qquad\qquad\qquad\qquad\qquad\qquad-\frac{\left|A\right|}{J^{2}}\left(R_{p}R'_{o}+R_{a}R_{p}+R_{a}R'_{o}\right)r+\frac{\left|A\right|}{J^{2}}R_{a}R_{p}R'_{o}.
\]
\\
It is important to recall that $A$ is negative for elliptic orbits, so it can be written as $A=-\left|A\right|$.

Comparing the coefficients of $r^{0}$, $r$ and $r^{2}$ of both sides of the last equation, and replacing the values of  $R_a$  and $R_p$ we obtain the following relations:

\begin{eqnarray}
\label{eq:Ro1}
\frac{r_{s}c^{2}}{J^{2}}=\frac{\left|A\right|}{J^{2}}\left(R'_{o}+2a\right),
\end{eqnarray}

\begin{eqnarray}
\label{eq:Ro2}
1= \frac{\left|A\right|}{J^{2}}\left(2aR'_{o}+a^{2}\left(1-e^{2}\right)\right)
\end{eqnarray}

\begin{eqnarray}
\label{eq:Ro3}
R'_{o}=\frac{J^{2}r_{s}}{\left|A\right|\left(1-e^{2}\right)a^{2}}\left(1-\frac{2\alpha}{cJ}E'\right)
\end{eqnarray}

The value of $R'_{0}$ given in (\ref{eq:Ro3}) is  the same result that we  obtained  in a previous paper \cite{MarinPoveda}, but with an extra factor $1-\frac{2\alpha}{cJ}E'$. Using equations (\ref{eq:Ro2}) and (\ref{eq:Ro3}) we can finally write:

\begin{eqnarray}
\label{eq:Rofinal}
R'_{0}=\frac{a\left(1-e^{2}\right)r_{s}\left(1-\frac{2\alpha E'}{cJ}\right)}{\left(a\left(1-e^{2}\right)-2r_{s}\left(1-\frac{2\alpha E'}{cJ}\right)\right)}, \label{eq:Roprime}
\end{eqnarray}

and

\begin{eqnarray}
\label{eq:JA}
\frac{J}{\left|A\right|^{\frac{1}{2}}}=\frac{a^{\frac{3}{2}}\left(1-e^{2}\right)}{\left(a\left(1-e^{2}\right)-2r_{s}\left(1-\frac{2\alpha E'}{cJ}\right)\right)^{\frac{1}{2}}}
\end{eqnarray}

Going back to equation (\ref{eq:kerr}), and because $\frac{dr}{d \tau} = \left(\frac{dr}{d \phi}\right)
\left(\frac{d \phi}{d \tau}\right)$, replacing the expression of $\frac{d \phi}{d \tau}$ given by (\ref{eq:dphitau}), we get:

\begin{eqnarray}
\left(\frac{dr}{d\phi}\right)^{2}\frac{1}{\gamma^2 c^2r^{4}}\left(\frac{r_{s}\alpha}{r}E'+\gamma cJ\right)^2=A+\frac{r_{s}c^{2}}{r}-\frac{J^{2}}{r^{2}}+\frac{J^{2}r_{s}}{r^{3}}-\frac{2r_{s}\alpha}{cr^{3}}JE'
\end{eqnarray}

Leaving only first order terms in  $\frac{\alpha }{r}$ we arrive to the following equation:

\begin{eqnarray}
\left(\frac{dr}{d\phi}\right)^{2}\left(1+\frac{2r_{s}\alpha E'}{\gamma rcJ}\right)=\frac{A}{J^{2}}r^{4}+\frac{r_{s}c^{2}}{J^{2}}r^{3}-r^{2}+\left(r_{s}-\frac{2r_{s}\alpha}{cJ}E'\right)r
\end{eqnarray}
that also can be written as:

\begin{eqnarray}
\left(\frac{dr}{d\phi}\right)^{2}\left(1+\frac{2r_{s}\alpha E'}{\gamma rcJ}\right)=\frac{\left|A\right|}{J^{2}}\left(R_{a}-r\right)\left(r-R_{p}\right)\left(r-R'_{o}\right)r.
\end{eqnarray}

The advance of the perihelion, then will be given by the integral:

\begin{eqnarray}
\Delta\phi_{kerr}=\frac{2J}{\left|A\right|^{1/2}}\intop_{R_{p}}^{R_{a}}\frac{\left(1+\frac{2r_{s}\alpha E'}{\gamma rcJ}\right)^{1/2}dr}{\left[\left(R_{a}-r\right)\left(r-R_{p}\right)\left(r-R'_{o}\right)r\right]^{1/2}}
\end{eqnarray}

and keeping only terms of up to first order in $\frac{\alpha}{r}$:

\begin{eqnarray}
\Delta\phi_{kerr}=\frac{2J}{\left|A\right|^{1/2}}\intop_{R_{p}}^{R_{a}}\frac{dr}{\left[\left(R_{a}-r\right)\left(r-R_{p}\right)\left(r-R'_{o}\right)r\right]^{1/2}}\qquad\qquad\qquad\qquad\qquad\qquad\
\nonumber
\end{eqnarray}

\begin{eqnarray}
\label{eq:kerr2}
\qquad\qquad\qquad\qquad\qquad\qquad\qquad+\frac{2r_s \alpha E'}{c\left|A\right|^{1/2}}\intop_{R_{p}}^{R_{a}}\frac{dr}{r\gamma\left[\left(R_{a}-r\right)\left(r-R_{p}\right)\left(r-R'_{o}\right)r\right]^{1/2}}
\end{eqnarray}

that can be written as:

\begin{eqnarray}
\Delta\phi_{kerr}=\Delta\phi_{sch}(R'_o)+\Delta\phi_{2}(R'_o)
\label{eq:Kerr3}
\end{eqnarray}
where:
\begin{eqnarray}
\Delta\phi_{sch}(R'_o)=\frac{2J}{\left|A\right|^{1/2}}\intop_{R_{p}}^{R_{a}}\frac{dr}{\left[\left(R_{a}-r\right)\left(r-R_{p}\right)\left(r-R'_{o}\right)r\right]^{1/2}},
\label{eq:kerr4}
\end{eqnarray}
and

\begin{eqnarray}
\Delta\phi_{2}(R'_o)=\frac{2r_s \alpha E'}{c\left|A\right|^{1/2}}\intop_{R_{p}}^{R_{a}}\frac{dr}{\left(r-r_s\right)\left[\left(R_{a}-r\right)\left(r-R_{p}\right)\left(r-R'_{o}\right)r\right]^{1/2}}
\label{eq:kerr5}
\end{eqnarray}

The first term in equation (\ref{eq:Kerr3})  is the same that was obtained in a previous paper \cite{MarinPoveda} using the Schwarzschild metric, but replacing $R_{o}$ by $R'_{o}$.

The parameter $\alpha$  is related to the spin of the black hole $s$, also known as the kerr parameter through the relationship:
\[
\alpha=\frac{GM}{c^{2}}s,
\]

in such a way that the angular momentum of spin is:

\begin{equation}
S_{z}=\frac{GM^{2}}{c}s,
\end{equation}

where $ s $ is a dimensionless parameter that can take values between 0 and 1. If $ s $ was greater than 1, there would be no event horizons and the singularity of $ r = 0 $ would be naked, which is not allowed \cite{Ryder,Wald}.

\section{Calculation of $\Delta\phi_{sch}(R'_o)$}

The first term in equation (\ref{eq:Kerr3}) also can be expressed as:

\begin{eqnarray}
\Delta\phi_{sch}(R'_o)=\frac{2J}{\left|A\right|^{1/2}}\intop_{R_{p}}^{R_{a}}\frac{r^{-\frac{1}{2}}\left(1-\frac{R'_{o}}{r}\right)^{-\frac{1}{2}}dr}{\left(\left(R_{a}-r\right)\left(r-R_{p}\right)r\right)^{\frac{1}{2}}},
\end{eqnarray}
and because $R'_{o}<<R_{p}<R_{a}$, we can expand around $R'_{o}$:

\begin{eqnarray}
\label{eq:R'oserie}
\left(1-\frac{R'_{0}}{r}\right)^{-\frac{1}{2}}=\sum_{n=1}^{\infty}\left(\begin{array}{c}
-\frac{1}{2}\\
n-1
\end{array} \right) \left(-1\right)^{n-1} \frac{\left(R'_{o}\right)^{n-1}}{r^{n-1}}.
\end{eqnarray}

Then
\begin{eqnarray}
\Delta\phi_{sch}(R'_o)=\frac{2J}{\left|A\right|^{1/2}}\sum_{n=1}^{\infty}\left(\begin{array}{c}
-\frac{1}{2}\\
n-1
\end{array} \right)\left(-1\right)^{n-1} \left(R'_{o}\right)^{n-1} I_{n},
\label{eq:Deltaphisch}
\end{eqnarray}
where:
\begin{eqnarray}
I_{n}=\intop_{R_{p}}^{R_{a}}\frac{dr}{r^{n}\left(\left(R_{a}-r\right)\left(r-R_{p}\right)\right)^{\frac{1}{2}}}.
\end{eqnarray}

The value of $I_{n}$ was calculated using complex integration in reference \cite{MarinPoveda}, where we obtained:

\begin{eqnarray}
I_{n}=\frac{\pi(-1)^{n+1}}{a^{n}2^{n-1}\left(1-e^{2}\right)^{n-1/2}}\sum_{k=0}^{n-1}\left(\begin{array}{c}
n-1\\
k
\end{array}\right)^{2}z_{1}^{n-1-k}z_{2}^{k}e^{n-1}
\label{eq:In}
\end{eqnarray}

where:

\begin{eqnarray}
z_{1}=-\frac{\left(1+\sqrt{1-e^{2}}\right)}{e},
\end{eqnarray}
and
\begin{eqnarray}
z_{2}=-\frac{\left(1-\sqrt{1-e^{2}}\right)}{e},
\end{eqnarray}

Defining the functions $Q_{n-1}\left(z_{1},z_{2}\right)$

\begin{eqnarray}
Q_{n}\left(z_1, z_2\right)=\sum_{k=0}^{n}\left(\begin{array}{c}
n\\
k
\end{array}\right)^{2}z_1^{n-k}{z_2}^{k}e^n,
\end{eqnarray}

we can write:

\begin{eqnarray}
\Delta\phi_{sch}(R'_o)=\frac{2 \pi J}{\left|A\right|^{1/2}}\sum_{n=0}^{\infty}\left(\begin{array}{c}
-\frac{1}{2}\\
n
\end{array}\right) \frac{\left(R'_{o}\right)^{n}Q_{n}\left(z_1, z_2\right)}{a^{n+1}2^{n}\left(1-e^{2}\right)^{\left(n+\frac{1}{2}\right)}}. 
\label{eq:Deltaphisch1}
\end{eqnarray}

In table (\ref{tab:AgujNegTabla-1}) it is shown the first five functions $Q_{n}$.
\begin{table}[h]
\caption{Values of the functions $Q_n$
 \label{tab:AgujNegTabla-1}}

\smallskip{}

\centering{}%
\begin{tabular}{ll}
\hline
\hline 
\noalign{\vskip\doublerulesep}
Function & Expression\tabularnewline[\doublerulesep]
\hline
\noalign{\vskip\doublerulesep}
$Q_{0}$ & $1$\tabularnewline[\doublerulesep]
\noalign{\vskip\doublerulesep}
$Q_{1}$ & $-2$\tabularnewline[\doublerulesep]
\noalign{\vskip\doublerulesep}
$Q_{2}$ & $\left(4+2e^{2}\right)$\tabularnewline[\doublerulesep]
\noalign{\vskip\doublerulesep}
$Q_{3}$ & $-\left(8+12e^{2}\right)$\tabularnewline[\doublerulesep]
\noalign{\vskip\doublerulesep}
$Q_{4}$ & $\left(16+48e^{2}+6e^{4}\right)$\tabularnewline[\doublerulesep]
\noalign{\vskip\doublerulesep}
$Q_{5}$ & $-\left(32+160e^{2}+60e^{4}\right)$\tabularnewline[\doublerulesep]
\hline
\hline
\end{tabular}
\end{table}

Introducing the values of $R'_{o}$ and  $\frac{J}{\left|A\right|^{1/2}}$  given by equations
(\ref{eq:Rofinal}) and  (\ref{eq:JA}), respectively, we get:

\begin{eqnarray}
\Delta\phi_{sch}(R'_o)=\frac{2 \pi}{\left(1-2 \epsilon \left(1- \frac{2 \alpha E'}{cJ}\right)\right)^{\frac{1}{2}}}\times \qquad\qquad\qquad\qquad\qquad \nonumber
\end{eqnarray}
\begin{eqnarray}
\label{eq:Deltaphi1}
 \qquad\qquad\qquad \times \sum_{n=0}^{\infty}\left(\begin{array}{c}
-\frac{1}{2}\\
n
\end{array}\right)\frac{ \left(1- \frac{2 \alpha E'}{cJ}\right)^{n}\epsilon^{n}}{2^{n}\left(1-2\epsilon\left(1- \frac{2 \alpha E'}{cJ}\right)\right)^{n}} Q_{n}\left(z_{1},z_{2}\right),
\end{eqnarray}
where
\begin{eqnarray}
\epsilon\equiv \frac{r_{s}}{a\left(1-e^{2}\right)}=\frac{2GM}{a\left(1-e^{2}\right)c^{2}},
\end{eqnarray}
and
\begin{eqnarray}
\left(\begin{array}{c}
-\frac{1}{2}\\
n
\end{array} \right)= \frac{\left(-1\right)^{n}\left(2n\right)!}{2^{2n}\left(n!\right)^{2}}.
\end{eqnarray}

\section{Calculation of $\Delta\phi_{2}(R'_o)$}

The integral $\Delta\phi_{2}(R'_o)$ also can be written as:
\begin{eqnarray}
\Delta\phi_{2}(R'_o)=\frac{2r_s \alpha E'}{c\left|A\right|^{1/2}}\intop_{R_{p}}^{R_{a}}\frac{\left(1-\frac{2r_{s}+R'_o}{r}+\frac{2R'_o r_{s}+r_{s}^{2}}{r^{2}}-\frac{r_{s}^{2}R'_o}{r^{3}}\right)^{-\frac{1}{2}}dr}{r^{2}\left[\left(R_{a}-r\right)\left(r-R_{p}\right)\right]^{1/2}}
\label{eq:Deltaphi2}
\end{eqnarray}

Let us call: $a_{1}=2r_{s}+R'_o$, $a_{2}=2R'_o r_{s}+r_{s}^{2}$, and $a_{3}=r_{s}^{2} R'_o$.
Performing the expansion:

\begin{eqnarray}
\left(1-\frac{a_{1}}{r}+\frac{a_{2}}{r^{2}}-\frac{a_{3}}{r^{3}}\right)^{-\frac{1}{2}}=\sum_{n=0}^{\infty}
\left(\begin{array}{c}
-\frac{1}{2}\\
n
\end{array}\right)\left(-\frac{a_{1}}{r}+\frac{a_{2}}{r^{2}}-\frac{a_{3}}{r^{3}}\right)^{n}=\sum_{n=0}^{\infty}\frac{c_{n}}{r^{n}},
\end{eqnarray}
we can check that for example:
\begin{eqnarray}
c_{0}=1,
\end{eqnarray}
\begin{eqnarray}
c_{1}=\frac{a_{1}}{2},
\end{eqnarray}
\begin{eqnarray}
c_{2}=\frac{1}{2}\left(\frac{3}{4} a_{1}^{2}-a_{2}\right),
\end{eqnarray}
\begin{eqnarray}
c_{3}=\frac{5}{16}a_{1}^{3}-\frac{3}{4}a_{1}a_{2}+\frac{1}{2}a_{3},
\end{eqnarray}
\begin{eqnarray}
c_{4}=\frac{35}{128}a_{1}^{4}+\frac{3}{4}a_{1}a_{3}-\frac{15}{16}a_{1}^{2}a_{2}+\frac{3}{8}a_{2}^{2},
\end{eqnarray}
\begin{eqnarray}
etc.... \nonumber
\end{eqnarray}

In general the $c_{n}$ satisfy the following recurrence relationship:
\begin{eqnarray}
c_{n+3}=\frac{1}{6+2n}\left(a_{1}c_{n+2}\left(5+2n\right)-2a_{2}c_{n+1}\left(2+n\right)+a_{3}c_{n}\left(3+2n\right)\right)
\end{eqnarray}

In terms of the $c_{n}$ we can write (\ref{eq:Deltaphi2}) as:
\begin{eqnarray}
\qquad\qquad \Delta\phi_{2}(R'_o)=\frac{2r_s \alpha E'}{c\left|A\right|^{1/2}}\intop_{R_{p}}^{R_{a}}\sum_{n=0}^{\infty}\frac{c_{n} dr}{r^{n+2}\left[\left(R_{a}-r\right)\left(r-R_{p}\right)\right]^{1/2}} \nonumber
\end{eqnarray}
\begin{eqnarray}
=\frac{2r_s \alpha E'}{c\left|A\right|^{1/2}}\sum_{n=0}^{\infty}c_{n}I_{n+2} \qquad
\end{eqnarray}

Replacing the value on $I_{n+2}$ given by (\ref{eq:In}) and the value of $\frac{J}{\left|A\right|^{1/2}}$ given by (\ref{eq:JA}) we get:
\begin{eqnarray}
\Delta\phi_{2}(R'_o)=\frac{\pi \epsilon}{2\left(1-2 \epsilon \left(1- \frac{2 \alpha E'}{cJ}\right) \right)^{\frac{1}{2}}}\left(\frac{2 \alpha E'}{cJ}\right)\sum_{n=0}^{\infty}\frac{\left(-1\right)^{n+1}c_{n}Q_{n+1}\left(z_{1},z_{2}\right)\epsilon^{n}}{2^{n}r_{s}^{n}}
\label{eq:deltaphi2Q}
\end{eqnarray}

\section{Expansion of $\Delta\phi_{sch}(R'_o)$ in terms of $\epsilon^{*} \equiv \left(1- \frac{2 \alpha E'}{cJ}\right) \epsilon $}

In terms of $\epsilon^{*} \equiv \left(1- \frac{2 \alpha E'}{cJ}\right) \epsilon $ (\ref{eq:Deltaphi1})
is:

\begin{eqnarray}
\Delta\phi_{sch}(R'_o)=\frac{2 \pi}{\left(1-2 \epsilon^{*} \right)^{\frac{1}{2}}} \
 \times \sum_{n=0}^{\infty}\left(\begin{array}{c}
-\frac{1}{2}\\
n
\end{array}\right)\frac{\left(\epsilon^{*}\right)^{n}}{2^{n}\left(1-2\epsilon^{*}\right)^{n}} Q_{n}\left(z_{1},z_{2}\right).
\label{eq:Kerrsch}
\end{eqnarray}

We will expand $\Delta\phi_{sch}(R'_o)$ in terms of $\epsilon^{*}$. Lets compute the first four terms of (\ref{eq:Kerrsch}) to recover the expansion until third order on $\epsilon^{*}$.

\begin{eqnarray}
\Delta\phi_{sch}(R'_o)^{(3)}=
\frac{2\pi}{\left(1-2\epsilon^{*}\right)^{1/2}}(Q_{0}\left(z_{1},z_2\right)- \left(\frac{1}{2}\right)\frac{Q_{1}\left(z_1,z_2\right)}{2\left(1-2\epsilon^{*}\right)}\epsilon^{*} + \left(\frac{3}{8}\right)\frac{Q_{2}\left(z_{1},z_2\right)}{2^2\left(1-2\epsilon^{*}\right)^2}\left(\epsilon^{*}\right)^2 \nonumber 
\end{eqnarray}
\begin{eqnarray}
-\left(\frac{5}{16}\right)\frac{Q_{3}\left(z_{1},z_2\right)}{2^3\left(1-2\epsilon^{*}\right)^3}\left(\epsilon^{*}\right)^3+ ......).
\end{eqnarray}  \\

Replacing the values of the $Q_{i}$ ($i=0,1,2,3$) we have:
\begin{eqnarray}
\Delta\phi_{sch}(R'_o)^{(3)}=
\frac{2\pi}{\left(1-2\epsilon^{*}\right)^{1/2}}(1+\frac{\epsilon^{*}}{2\left(1-2\epsilon^{*}\right)}+
\left(\frac{3}{16}\right)\frac{\left(2+e^{2}\right)\left(\epsilon^{*}\right)^{2}}{\left(1-2\epsilon^{*}\right)^{2}} \nonumber
\end{eqnarray}
\begin{eqnarray}
+\left(\frac{5}{32}\right)\frac{\left(2+3e^{2}\right)\left(\epsilon^{*}\right)^{3}}{\left(1-2\epsilon^{*}\right)^{3}} + ........)
\label{eq:phisch2} 
\end{eqnarray}
which in third orden in $\epsilon^{*}$ is reduced to:
\begin{eqnarray}
\Delta\phi_{sch}(R'_o)^{(3)}=2\pi\left(1+\frac{3}{2}\epsilon^{*}+\frac{\left(54+3e^{2}\right)}{16}
\left(\epsilon^{*}\right)^{2}+\left(\frac{135}{16}+\frac{45}{32}e^{2}\right)\left(\epsilon^{*}\right)^{3}+ ....\right).
\end{eqnarray}

In terms of $\epsilon$ we have:
\begin{eqnarray}
\Delta\phi_{sch}(R'_o)^{(3)}=2\pi (1+\frac{3}{2}\left(1- \frac{2 \alpha E'}{cJ}\right) \epsilon + \frac{3\left(18+e^{2}\right)}{16}
\left(1- \frac{2 \alpha E'}{cJ}\right)^{2} \epsilon^{2} \nonumber
\end{eqnarray}
\begin{eqnarray}
+\frac{45}{32}\left(6+e^{2}\right)\left(1- \frac{2 \alpha E'}{cJ}\right)^{3} \epsilon^{3}+ ....).
\end{eqnarray}

In general to order $m$ in $\epsilon^{*}$ we can write:
\begin{eqnarray}
\Delta\phi_{sch}(R'_o)^{(m)}=\frac{2 \pi}{\left(1-2 \epsilon^{*} \right)^{\frac{1}{2}}} \
 \times \sum_{n=0}^{m}\left(\begin{array}{c}
-\frac{1}{2}\\
n
\end{array}\right)\frac{\left(\epsilon^{*}\right)^{n}}{2^{n}\left(1-2\epsilon^{*}\right)^{n}} Q_{n}\left(z_{1},z_{2}\right).
\label{eq:deltaphim}
\end{eqnarray}

\section{Expansion of $\Delta\phi_{2}(R'_o)$ in terms of $ \epsilon $}
To evaluate $\Delta\phi_{2}(R'_o)$ to third order in $ \epsilon $ it is sufficient to consider the first three terms in the sum given by  equation (\ref{eq:deltaphi2Q}).
\begin{eqnarray}
\Delta\phi_{2}(R'_o)^{\left(3\right)}=\frac{\pi \epsilon}{2\left(1-2 \epsilon \left(1- \frac{2 \alpha E'}{cJ}\right) \right)^{\frac{1}{2}}}\left(\frac{2 \alpha E'}{cJ}\right)\left(-c_{0}Q_{1}+\frac{c_{1}Q_{2}\epsilon}{2r_{s}}-\frac{c_{2}Q_{3}\epsilon^{2}}{2^{2}r_{s}^{2}}\right),
\label{eq:deltaphi*}
\end{eqnarray}
where 
\begin{equation}
c_{0}=1,
\end{equation}
\begin{equation}
c_{1}=\frac{2r_{s}+R'_o}{2},
\end{equation}
\begin{equation}
c_{2}=\frac{3}{8}\left(R'_o\right)^{2}+\frac{1}{2}r_{s}R'_o+r_{s}^{2}.
\end{equation}

If we introduce the values of the $Q_{i}$ in (\ref{eq:deltaphi*}) we have:
\begin{eqnarray}
\Delta\phi_{2}(R'_o)^{\left(3\right)}=\frac{\pi \epsilon}{\left(1-2 \epsilon \left(1- \frac{2 \alpha E'}{cJ}\right) \right)^{\frac{1}{2}}}\left(\frac{2 \alpha E'}{cJ}\right)(1+\frac{\left(2r_{s}+R'_o\right)\left(2+e^{2}\right)\epsilon}{4r_{s}} \nonumber
\end{eqnarray}
\begin{eqnarray}
+\frac{\left(3\left(R'_o\right)^{2}+4r_{s}R'_o+8r_{s}^{2}\right)\left(2+3e^{2}\right)\epsilon^{2}}{16r_{s}^{2}}),
\label{eq:deltaphi24}
\end{eqnarray}
where 
\begin{eqnarray}
\frac{\left(2r_{s}+R'_o\right)}{r_{s}}=\frac{2+\left(1-4\epsilon \right) \left(1- \frac{2 \alpha E'}{cJ}\right)}{\left(1-2\epsilon \left(1- \frac{2 \alpha E'}{cJ}\right)\right)},
\label{eq:par1}
\end{eqnarray}
and
\begin{eqnarray}
\frac{\left(3\left(R'_o\right)^{2}+4r_{s}R'_o+8r_{s}^{2}\right)}{r_{s}^{2}}=\frac{8+4\left(1- \frac{2 \alpha E'}{cJ}\right)\left(1-8\epsilon\right)+\left(1- \frac{2 \alpha E'}{cJ}\right)^{2}\left(3-8\epsilon+32\epsilon^{2}\right)}{\left(1-2\epsilon \left(1- \frac{2 \alpha E'}{cJ}\right)\right)^{2}}.
\label{eq:par2}
\end{eqnarray}

Introducing (\ref{eq:par1}) and (\ref{eq:par2}) in (\ref{eq:deltaphi24}), and employing the Taylor series (around $x=0$): $\left(1-x\right)^{-\frac{1}{2}}=1+\frac{1}{2}x+\frac{3}{8}x^{2}+...$, we obtain to third order in $\epsilon$:
\begin{eqnarray}
\Delta\phi_{2}(R'_o)^{\left(3\right)}=\pi \frac{2 \alpha E'}{cJ} \epsilon +\frac{\pi}{4}\left(10+3e^{2}-\left(6+e^{2}\right)\frac{2 \alpha E'}{cJ}\right)\frac{2 \alpha E'}{cJ}\epsilon^{2}
\nonumber
\end{eqnarray}
\begin{eqnarray}
+\frac{\pi}{16}\left(86+53e^{2}-\left(116+54e^{2}\right)\frac{2 \alpha E'}{cJ}+\left(38+13e^{2}\right)\left(\frac{2 \alpha E'}{cJ}\right)^{2}\right)\frac{2 \alpha E'}{cJ}\epsilon^{3}
\label{eq:deltaphi23}
\end{eqnarray}
\section{Perihelion precession to third order in $\epsilon$}
As the perihelion precession in a cycle is $\chi=\Delta\phi_{kerr}-2\pi$, we have until third order in $\epsilon$:
\begin{eqnarray}
\chi^{\left(3\right)}=\Delta\phi_{sch}(R'_o)^{(3)}+\Delta\phi_{2}(R'_o)^{\left(3\right)}-2\pi,
\end{eqnarray}
that give us:
\begin{eqnarray}
\chi^{\left(3\right)}=3\pi\left(1-\frac{4 \alpha E'}{3cJ}\right)\epsilon +\left(\frac{\left(54+3e^{2}\right)}{8}-11\left(\frac{2 \alpha E'}{cJ}\right)+\frac{\left(42+e^{2}\right)}{8}\left(\frac{2 \alpha E'}{cJ}\right)^{2}\right)\pi\epsilon^{2}
\nonumber
\end{eqnarray}
\begin{eqnarray}
+(\frac{45\left(6+e^{2}\right)}{16}-\frac{\left(362+41e^{2}\right)}{8}\left(\frac{2 \alpha E'}{cJ}\right)+\frac{\left(694+81\right)}{16}\left(\frac{2 \alpha E'}{cJ}\right)^{2}
\nonumber
\end{eqnarray}
\begin{eqnarray}
-\frac{\left(29+4e^{2}\right)}{2}\left(\frac{2 \alpha E'}{cJ}\right)^{3})\pi \epsilon^{3}.
\label{eq:chi3} 
\end{eqnarray}

\section{Applications}

To analyze how the spin of the central black hole changes the perihelion precession, it was calculated until 3rd order (using equation (\ref{eq:chi3})) for three different binary systems. In table (\ref{tab:Oj287}) are shown the results. The term $\chi(\epsilon^i)$ is the one that depends only on the i-th power of $\epsilon$, since $\chi^{(i)}=\chi(\epsilon^1)+\chi(\epsilon^2)+...+\chi(\epsilon^i)$. The expansion was done in terms of $\alpha$, that is a parameter with units of length, but the experimental measurements are done in terms of the spin $s$. The spin $s$ of the black hole is a dimensionless parameter that is related to $\alpha$ by:

\begin{equation}
\alpha=\frac{GM}{c^2}s=\frac{r_s}{2}s
\end{equation}

The spin of a central black hole in a binary system is very difficult to measure. Nevertheless, some groups have measured it for the binary system OJ287. This measurements can be seen in articles by Pihajoki and Valtonen \cite{Pihajoki,Valtonen}. The problem is that the values obtained form those measurements are different, however we worked with both values for the numerical evaluations. Also we made the calculations for $s=0$.

For the calculations we need the values of the energy per unit mass and the angular momentum per unit mass. For this, both values can be calculated numerically using equations (\ref{eq:3.4-1})
and (\ref{eq:3.5-1}). These are the equations of motion evaluated for the critical points $R_a$ (aphelion) and $R_p$ (perihelion). As the equations are quadratic, there are two solutions for each parameter. We took the negative value for the energy (that is necessary for the orbit to be elliptical), and the positive value for the angular momentum, because we considered a counterclockwise rotation.

\begin{table}[h]
\caption{Perihelion advance in degrees per period for some binary systems, and for different values of the spin. \label{tab:Oj287}}
\smallskip{}

\centering{}%
\begin{tabular}{llllll}
\hline
System & Sagittarius A*-S2  & OJ287 & H1821+643 \tabularnewline
\hline
$M(\times M_{\Theta})$ & $4.310\times10^{6}$ & $1.830\times10^{10}$ & $3.000\times10^{10}$ \tabularnewline
$r_s(AU)$ & $0.085$ & $360.847$ & $591.553$ \tabularnewline
$a(AU)$ & $923.077$ & $11500$ & $40000$ \tabularnewline
$e$ & $0.870$ & $0.700$ & $0.900$ \tabularnewline
$\epsilon$ & $3.787\times 10^{-4}$ & $6.153\times 10^{-2}$ & $7.784\times 10^{-2}$ \tabularnewline
\hline
$s$ & $0$ & $0$ & $0$ \tabularnewline
\hline
$\chi(\epsilon)$ & $0.205^\circ$ & $33.223^\circ$ & $42.031^\circ$ \tabularnewline
$\chi(\epsilon^2)$ & $(1.816\times 10^{-4})^\circ$ & $4.724^\circ$ & $7.692^\circ$ \tabularnewline
$\chi(\epsilon^3)$ & $(1.858\times 10^{-7})^\circ$ & $0.765^\circ$ & $1.625^\circ$ \tabularnewline
$\chi^{(3)}$ & $0.205^{\circ}$ & $38.713^{\circ}$ & $51.349^{\circ}$ \tabularnewline
\hline
$s$ & $0.280$ & $0.280$ & $0.280$ \tabularnewline
\hline
$\chi(\epsilon)$ & $0.206^\circ$ & $35.249^\circ$ & $46.230^\circ$ \tabularnewline
$\chi(\epsilon^2)$ & $(1.837\times 10^{-4})^\circ$ & $5.440^\circ$ & $9.622^\circ$ \tabularnewline
$\chi(\epsilon^3)$ & $(1.895\times 10^{-7})^\circ$ & $0.965^\circ$ & $2.349^\circ$ \tabularnewline
$\chi^{(3)}$ & $0.206^{\circ}$ & $41.654^{\circ}$ & $58.203^{\circ}$ \tabularnewline
\hline
$s$ & $0.313$ & $0.313$ & $0.313$ \tabularnewline
\hline
$\chi(\epsilon)$ & $0.206^\circ$ & $35.488^\circ$ & $46.727^\circ$ \tabularnewline
$\chi(\epsilon^2)$ & $(1.841\times 10^{-4})^\circ$ & $5.528^\circ$ & $9.866^\circ$ \tabularnewline
$\chi(\epsilon^3)$ & $(1.899\times 10^{-7})^\circ$ & $0.991^\circ$ & $2.448^\circ$ \tabularnewline
$\chi^{(3)}$ & $0.206^{\circ}$ & $42.007^{\circ}$ & $59.420^{\circ}$ \tabularnewline
\hline 
\end{tabular}
\end{table}

In table (\ref{tab:Oj287}) it can be seen the values for $s=0$ for the systems Sagittarius A*-S2, OJ287, H1821+643. These values agree with those calculated in a previous paper \cite{MarinPoveda} using the Schwarzschild metric. These results are in concordance with the fact that the Kerr space-time transforms in Schwarzschild space-time when $s=0$.

Then, for other values of $s$, the perihelion precession increases with $s$. Also, it is important to recall that we considered that $s$ is positive, i.e, the central black hole rotates in the same direction of the movement of the secondary massive object. We have used the values of the spin of OJ287 with all the systems to see how the perihelion precession changes with it because we do not have information about the values for those systems.

For OJ287, the experimental accepted value for the perihelion precession is $39.1^\circ$ per period. The theoretical calculation (for $s=0.313$) predicts a higher value of $42.0^\circ$. Without considering the spin, the theoretical value is $38.7^{\circ}$. It was expected that the introduction of the spin would reduce the error with respect to the experimental value, but instead it exceeded this value. Nevertheless, there is another correction that is expected to reduce the value of the perihelion precession. This correction consist in taking into account the lost of energy in the gravitational radiation. That correction will be done in further works.

\section{Conclusions}
\label{Conclusions}

In this paper we have considered both Schwarzschild and Kerr metrics to take into account the spin effect in the calculation of the perihelion precession in binary systems like OJ287, Sagittarius A*-S2, and H1821+643. We have developed the expansion in terms of both $\epsilon \equiv r_s/\left(a(1-e^2)\right)$ and  $\epsilon^{*} \equiv \left(1- \frac{2 \alpha E'}{cJ}\right) \epsilon $, where $E'$ and $J$ are the energy and angular momentum per unit mass and $\alpha$ is a factor proportional to the black hole spin. Here we have used the notation $\chi\left(\epsilon^{n}\right)$ for the contribution of the $n^{th}$ term and $\chi^{\left(n\right)}$  for the complete expansion until $n^{th}$ term.

Now, to see if the spin correction is relevant for  massive objects, we calculated the perihelion advance for three binary systems, Sagittarius A*-S2, OJ287 and H1821+643. In table (\ref{tab:Oj287}) are shown these calculations performed using equation (\ref{eq:chi3})   for different orders.

Sagitarius A* is a bright and very compact radio source located at the center of the Milky Way. There is great evidence that Sagittarius A* is a supermassive black hole wtith a mass approximately equal to $4	\times 10^{6} M_{\Theta}$ . We took star S2, because is the one that presents a very peculiar orbit. As it can be seen in table (\ref{tab:Oj287}), if we compare the value of $\chi^{(3)}$ for $s=0$ that is $0.205^\circ$ with the corresponding values for $s=0.280$ and $s=0.313$ that is $0.206^\circ$, we can see that the difference is not significant. It is worth mentioning that the values of $\chi\left(\epsilon^{n}\right)$ and  $\chi^{\left(n\right)}$ corresponding to $s=0$ are agree with those calculated in a
earlier paper \cite{MarinPoveda} using the Schwarzschild metric.

OJ 287 is a binary system of black holes located 3.500 million light years from Earth having a total mass of around $1.845\times 10^{10}M_{\Theta}$. It can be seen that between the first and the second order terms ($\chi\left(\epsilon\right)$ and $\chi\left(\epsilon^{2}\right)$)  there is a difference of approximately $5.5^{\circ}$. For higher orders, the difference is less than $1^{\circ}$. At the third order the contribution to  the perihelion precession in a cycle  is $0.965^{\circ}$ in our expansion for $s=0.280$  compared with the value of $0.765^{\circ}$ obtained in a earlier paper \cite{MarinPoveda} taking $s=0$. Then, the total perihelion precession is $41.654^{\circ}$. For $s=0.313$ the perihelion precession is slightly different (approximately $42^{\circ}$).  Something interesting about our expansion is that it begins to stabilize taking into account higher order corrections around $42^{\circ}$.  Then, equation (\ref{eq:chi3}))  gives an important correction to the perihelion precession. It is important to mention that  the gravitational radiation of the system (that affects directly to the values of the period, the radial distance and the eccentricity of the orbit \cite{Shreya}), could introduce important modifications in the perihelion advance equation. 

For H1821+643 system , that is a very massive black hole with a mass of $3\times 10^{10}M_{\Theta}$, the orbital parameters of the gravitational companion are not known, so we used random parameters. In this system, the corrections are very important. For example, taking $s=0.313$, the second order contribution $\chi\left(\epsilon^{2}\right)$ is $9.866^{\circ}$, while the third order contribution $\chi\left(\epsilon^{3}\right)$ is $2.448^{\circ}$ .

Taking into account all these results, we can conclude that this is an excellent approach to calculate the perihelion precession for binary systems as it can be calculated until any order in the third root of the motion equation $R'_o$ (see equation (\ref{eq:Roprime})). With other methods as perturbation theory, the calculations are more complicated and due to the approximations that are made it is easy to make mistakes.

\end{document}